\documentclass[12pt,epsf]{article}
\usepackage{epsfig}
\usepackage{amsmath}
\usepackage{amssymb}
\usepackage{graphicx}
\newcommand{\eps}{\epsilon}

\newcommand{\up}{\uparrow}
\newcommand{\down}{\downarrow}
\newcommand{\updown}{\updownarrow}

\newcommand{\be}{\begin{equation}}
\newcommand{\ee}{\end{equation}}
\newcommand{\ben}{\begin{eqnarray}\displaystyle}
\newcommand{\een}{\end{eqnarray}}

\newcommand{\ta}{\tilde\alpha}

\newcommand{\s}{\sigma}

\newcommand{\sectiono}[1]{\section{#1}\setcounter{equation}{0}}
\newcommand{\brangle}{{\big\rangle}}
\renewcommand{\theequation}{\thesection.\arabic{equation}}

\begin{document}

{}~ \hfill\vbox{
\hbox{HUTP-06/A0008}}\break

\vskip 1.cm

\centerline{\large \bf Strong coupling from the Hubbard model}
\vspace*{5.0ex}

\centerline{\large \rm Joseph~A.~Minahan\footnote{On leave from Institutionen f\"or Teoretisk Fysik, Uppsala Universitet, Uppsala, Sweden}}

\vspace*{2.5ex}
\centerline{\large \it Jefferson Physical Laboratory }
\centerline{\large \it Harvard University, Cambridge, MA 02138 USA}
\vspace*{2.5ex}

\centerline{\tt joseph.minahan@teorfys.uu.se}

\vspace*{1.0ex}
\medskip
\bigskip\bigskip
\centerline {\bf\large Abstract}

\bigskip\bigskip

It was recently observed that the one dimensional half-filled Hubbard model  reproduces the known part of the perturbative  spectrum of planar ${\cal N}=4$ super Yang-Mills in the $SU(2)$ sector.  Assuming that this identification is valid beyond perturbation theory, we investigate the behavior of this spectrum as the 't Hooft parameter $\lambda$ becomes large.  We show that the full dimension $\Delta$ of the Konishi superpartner is the solution of a sixth order polynomial while $\Delta$ for a bare dimension 5 operator is the solution of a cubic.   In both cases the equations  can be solved easily as a series expansion for both small and large $\lambda$ and the equations can be inverted to express $\lambda$  as an explicit function of $\Delta$.  We then consider more general operators and show how $\Delta$ depends on $\lambda$ in the strong coupling limit.  We are also able to distinguish those states in the Hubbard model which correspond to the gauge invariant operators for all values of $\lambda$.  Finally, we compare our results with  known results for strings on $AdS_5\times S^5$, where we find agreement for a range of $R$-charges.

\bigskip

\vfill \eject
\baselineskip=17pt

\sectiono{Introduction}

The underlying integrability of planar ${\cal N}=4$ super Yang-Mills is expected to be crucial in  better understanding the AdS/CFT correspondence \cite{Maldacena:1997re,Gubser:1998bc,Witten:1998qj}.  At one loop integrability is well-established, since the gauge invariant operators can be mapped to the states of an integrable spin chain \cite{Minahan:2002ve,Beisert:2003yb}.  In this map the dilatation operator plays the role of the spin chain Hamiltonian.  But higher loop integrability requires the inclusion of all loops, making it difficult to prove. Nevertheless, important progress has been made in establishing higher loop integrability, starting with the work in \cite{Beisert:2003tq}, and many further tests have been carried out \cite{Beisert:2003ys,Beisert:2004ry}.  The dilatation operator is known with some amount of confidence up to five loops in the reduced $SU(2|3)$ sector but an all loop operator seems out of reach, or if it could be reached, not particularly useful.  

On the other hand, it might not be necessary to find the dilatation operator if one knows the  2-body $S$-matrix and a dispersion relation for the ``magnons" on the spin chain.  With the $S$-matrix, one can construct Bethe ansatz equations for the magnon momenta and from there compute the energies of the states \cite{Staudacher:2004tk,Beisert:2005fw}.   In \cite{Beisert:2004hm}, Beisert Dipple and Staudaucher (BDS) gave a conjecture  for the asymptotic Bethe ansatz in the $SU(2)$ sector which is consistent with the three loop dilatation operator and which has a form similar to an integrable equation derived from classical string motion on $R\times S^3$ \cite{Kazakov:2004qf} .   Unlike a previous conjecture \cite{Serban:2004jf},  the BDS Bethe equations are consistent with all loop BMN scaling \cite{Berenstein:2002jq}.  Using the BDS equations it is possible to compute the largest energy state on the chain \cite{Rej:2005qt,Zarembo:2005ur}, where it was discovered by Rej, Serban and Staudacher (RSS) that it had precisely the same energy as a Lieb-Wu state for the one-dimensional half-filled Hubbard model \cite{Lieb_Wu}, with a hopping parameter dependent on the 't Hooft parameter $\lambda$ \cite{Rej:2005qt}.  RSS were also able to show that integrating out the hopping terms for the half filled states of the Hubbard model led to an effective Hamiltonian that matched the 3 loop dilatation operator.  They further showed that the magnon dispersion relation and the asymptotic $S$-matrix for two magnons reproduces the BDS result.  Finally, RSS demonstrated that the Hubbard model's Bethe equations could be used to find a series expansion for operators with small bare dimensions, including the dimension 4 operator that is part of the Konishi multiplet.  

It is not presently clear why the Hubbard model should reproduce perturbation theory.  This is an important question that hopefully can be answered.  In the meantime, one can treat the RSS identification of gauge invariant operators with a subset of Hubbard model states  as an ansatz, even at strong coupling, and explore its consequences.  

 While Bethe ansatz equations are in principle solvable, actually finding the solutions is often cumbersome.  Even the 4 and 5 site series solutions in \cite{Rej:2005qt} are quite involved.   However, the Hubbard model has a simple Hamiltonian, and since it only acts on a finite system of states, it is possible to find the desired energies by solving a polynomial equation.  For example, there are 36 half-filled states with an equal number of up and down spins on the 4 site Hubbard model.  But invoking symmetries reduces the number of relevant states to 6.   On the half-filled 5 site Hubbard model with three up and two down spins there are 100 states, but symmetries reduce these to nine.  However, it turns out that the resulting ninth order polynomial is itself factorizable into a cubic and a sextic polynomial, where the solution corresponding to the operator in the gauge theory is a zero of the cubic.

A key point about the Hubbard model is that for large $\lambda$ it reduces to a theory of free lattice fermions.  However, one still needs to determine how the weakly coupled gauge theory states flow to the lattice fermion states as the coupling is increased.  We will show how to do this explicitly for a particular class of states using the nested Bethe ansatz and sketch the argument for more general states.
Along the way we will give a criteria that identifies which Hubbard model states are to be identified with gauge invariant operators.  

Once we are in the strong coupling region where $\lambda>>1$, we can compare the results to known results in string theory.  In particular, we will consider the case of operators with bare dimension $L$ and $R$-charge $L-2$, but which are outside the BMN scaling range \cite{Berenstein:2002jq}.  We will show that there is  a range of $R$-charges where the results agree with a prediction in \cite{Gubser:1998bc} for the dimension of these operators.

In section 2 we consider specific solutions for Hubbard chains with 4 and 5 sites respectively.  We identify the states that correspond to gauge invariant operators and show that $\Delta$ is a solution to a 
polynomial equation.  In section 3 we consider more general operators and compute $\Delta$ for these operators in the limit of large 't Hooft coupling $\lambda$.  In section 4 we compare our results from section 3 with string theory results.

\sectiono{Solutions for 4 and 5 sites}

In this section we compute the dimensions for operators which  are not chiral primaries, and that have bare dimensions 4 and 5, using the Hubbard model ansatz.
The Hamiltonian for the Hubbard model which reproduces the BDS Bethe ansatz is given \mbox{by \cite{Rej:2005qt}}
\begin{equation}\label{ham}
H=\frac{g}{\sqrt{2}}\Bigg(\sum_{i=1}^{L-1}\Big(c^\dagger_{\s,i} c_{\s,i+1}+c_{\s,i+1}^\dagger c_{\s,i}\Big)+e^{i\phi}c_{\s,L}^\dagger c_{\s,1}+e^{-i\phi}c_{\s,1}^\dagger c_{\s,L}\Bigg)+\sum_{i=1}^L(1-c_{\up,i}^\dagger c_{\up,i}c_{\down,i}^\dagger c_{\down,i})
\end{equation}
where $\phi=\pi/2\ (0)$ if $L$ is even (odd).  For what follows we will leave the phase $\phi$ arbitrary, eventually setting it to the values used in \cite{Rej:2005qt}.  The parameter $g$ is given by 
\begin{equation}
g^2=\frac{(g_{YM})^2N}{8\pi^2}=\frac{\lambda}{8\pi^2}\,.
\end{equation}
We have chosen to normalize the Hamiltonian as in \cite{Roiban:2006jt} so that the energy is given by the bare dimension when $g=0$.   The parameter $\phi$ arises from an Aharanov-Bohm flux which is needed to reproduce the BDS results \cite{Rej:2005qt}.   The last term in (\ref{ham}) is an interaction term between the lattice fermions which lowers the energy if two fermions are occupying the same site.  The states corresponding to the gauge invariant operators are not the lowest energy states \cite{Rej:2005qt}.  The Hamiltonian is also invariant under the shift symmetry
\begin{eqnarray}\label{shift}
&&c_{\s,j}\to e^{i\phi /L}c_{\s,j+1},\qquad c_{\s,j}^{\dag}\to e^{-i\phi /L}c_{\s,j+1}^{\dag},\qquad\qquad   j<L\nonumber\\ &&c_{\s,L}\to e^{i\phi (L+1)/L}c_{\s,1}\qquad c_{\s,L}^{\dag}\to e^{-i\phi(L+1)/L}c^{\dag}_{\s,1}\,.
\end{eqnarray}

Let us first consider the Hubbard model with just 4 sites.  There are 36 possible half-filled states and the Hamiltonian can mix many of these states amongst each other.  However, the symmetries of the system will reduce the mixing to a smaller subset of states.   Since the total spin is conserved, we can reduce our space to states with total  spin zero.  We can further reduce the states to those that are invariant under the shift in (\ref{shift}).  
A basis for these states is given by
\begin{eqnarray}\label{basis}
\big| A\brangle &= &\frac{1}{2\sqrt{3}}\Big(2\big| \up\down\up\down\brangle+2\big| \down\up\down\up\brangle
-\big| \up\up\down\down\brangle-\big| \up\down\down\up\brangle-\big| \down\down\up\up\brangle-\big| \down\up\up\down\brangle\Big)\nonumber\\
\big| B_1\brangle&=&\frac{1}{2\sqrt{2}}\Big(\big| \updown0\up\down\brangle+\big| \down\updown0\up\brangle+\big| \up\down\updown0\brangle-e^{i\phi}\big| 0\up\down\updown\brangle
\nonumber\\
&&\qquad-\big| \updown0\down\up\brangle-\big| \up\updown0\down\brangle-\big| \down\up\updown0\brangle+e^{i\phi}\big| 0\down\up\updown\brangle\Big)\nonumber\\
\big| B_2\brangle&=&\frac{1}{2\sqrt{2}}\Big(e^{-i\phi/2}\big| \updown\up0\down\brangle+e^{-i\phi/2}\big| \down\updown\up0\brangle-e^{i\phi/2}\big| 0\down\updown\up\brangle-e^{i\phi/2}\big| \up0\down\updown\brangle
\nonumber\\
&&\qquad-e^{-i\phi/2}\big| \updown\down0\up\brangle-e^{-i\phi/2}\big| \up\updown\down0\brangle+e^{i\phi/2}\big| 0\up\updown\down\brangle+e^{i\phi/2}\big| \down0\up\updown\brangle\Big)\nonumber\\
\big| B_3\brangle&=&\frac{1}{2\sqrt{2}}\Big(\big| 0\updown\up\down\brangle+\big| \down0\updown\up\brangle+\big| \up\down0\updown\brangle-e^{-i\phi}\big| \updown\up\down0\brangle
\\
&&\qquad-\big| 0\updown\down\up\brangle-\big| \up0\updown\down\brangle-\big| \down\up0\updown\brangle+e^{-i\phi}\big| \updown\down\up0\brangle\Big)\nonumber
\\ 
\big| C_1\brangle&=&\frac{1}{\sqrt{2}}\Big(e^{-i\phi/2}\big| \updown0\updown0\brangle-e^{i\phi/2}\big| 0\updown0\updown\brangle\Big)\nonumber\\
\big| C_2\brangle&=&\frac12\Big(\big| \updown00\updown\brangle-e^{-i\phi}\big| \updown\updown00\brangle+\big| 0\updown\updown0\brangle-e^{i\phi}\big| 00\updown\updown\brangle\Big)\nonumber
\end{eqnarray}
The states are defined by the fermion occupancy at each site, with the fermion creation operators ordered by increasing lattice number and with double occupied sites ordered with the down creation operator to the right of the up.  So for example 
\begin{equation}\big| \up\down0\updown\brangle\equiv c_{\up,1 }^{\dagger}c_{\down,2}^{\dagger}c_{\up,4}^{\dagger}c_{\down,4 }^{\dagger}\big| 0\brangle.
\end{equation}
Using the basis of states in (\ref{basis}) it is straightforward to show that $H$ in (\ref{ham}) is given by
\begin{equation}
H=\left(\begin{array}{cccccc}
4&\sqrt{3}g&0&\sqrt{3}g&0&0\\
\sqrt{3}g&3&e^{i\phi/2}\sqrt{2}g&0&e^{i\phi/2}\sqrt{2}g&g\\
0&e^{-i\phi/2}\sqrt{2}g&3&-e^{i\phi/2}\sqrt{2}g&0&0\\
\sqrt{3}g&0&-e^{-i\phi/2}\sqrt{2}g&3&-e^{-i\phi/2}\sqrt{2}g&g\\
0&e^{-i\phi/2}\sqrt{2}g&0&-e^{i\phi/2}\sqrt{2}g&2&0\\
0&g&0&g&0&2
\end{array}\right)
\end{equation}
Hence, the secular equation giving the eigenvalues is a sixth order polynomial and is given by
\begin{eqnarray}\label{secular}
\det(\Delta-H)&=&0\nonumber\\
&=&\Delta^6-17\Delta^5+(119-16g^2)\Delta^4-(439-176g^2)\Delta^3\nonumber\\
&&+(900-716g^2+64\cos^2\frac\phi2\,g^4)\Delta^2
-(972+1276g^2-320\cos^2\frac\phi2\,g^4)\Delta\nonumber\\
&&+432-840g^2+400\cos^2\frac\phi2\,g^4\,.
\end{eqnarray}
One can easily solve this as a series expansion, where the relevant solution is given by
\begin{eqnarray}
\Delta&=&4 + 6\ g^2 -6(2+3\cos\phi)\ g^4 + 6(7+45\cos\phi)\ g^6 \nonumber\\
&&-6(53+486\cos\phi+144\cos^2\phi)\ g^8 + 6(754+4515\cos\phi+4815\cos^2\phi)\  g^{10} \nonumber\\
&&-6(10631+42597\cos\phi+94905\cos^2\phi+11637\cos^3\phi)\ g^{12}\nonumber\\
&& +12(65327+243243\cos\phi+718704\cos^2\phi+299754\cos^3\phi)\ g^{14} +\dots\ \,,
\end{eqnarray}
which matches the expansion found in \cite{Rej:2005qt} after setting $\phi=\pi/2$.

Since the polynomial in (\ref{secular}) is quadratic in $g^2$, we can write a simple closed form solution for $\lambda$  in terms of $\Delta$, 
\begin{eqnarray}\label{lambdaeq}
\lambda& =&\frac{\pi^2}{\cos^2\phi/2}\frac{(\Delta-2)(\Delta-3)}{2\Delta-5}\left(2\Delta-7-\sqrt{(2\Delta-7)^2\sin^2\frac{\phi}{2}+\cos^2\frac{\phi}{2}}\right)\,,
\end{eqnarray}
which after setting $\phi=\pi/2$ can be expressed in terms of the dimension of the Konishi operator, $\Delta_K=\Delta-2$,  as
\begin{eqnarray}\label{lambdaeqK}
\lambda& =&2\pi^2\ \frac{\Delta_K(\Delta_K-1)}{2\Delta_K-1}\left(2\Delta_K-3-\sqrt{2\Delta_K^2-6\Delta_K+5}\right)
\,.
\end{eqnarray}
{}From the expression in (\ref{lambdaeq}) it is easy to see that the behavior of $\Delta$ for large $\lambda$ is
\begin{equation}\label{limit}
\Delta\approx \frac{\sqrt{1+\sin\phi/2}}{\pi}\,\sqrt{\lambda}\,.
\end{equation}
A similar expression with a slightly higher coefficient  was found in \cite{Ryzhov:2004nz,Tseytlin:2004xa}, where the dilatation operator was reduced to the linear part of the spin-spin interaction.

We can also see the behavior in (\ref{limit}) arise from the original Hamiltonian in (\ref{ham}), where in the limit of large coupling,
the hopping term will dominate over the contact term.  In this case the theory reduces to free lattice fermions with a dispersion relation \cite{Rej:2005qt}
\begin{equation}\label{dispersion}
\eps(k)=\frac{\sqrt{\lambda}}{2\pi}\,\cos(k+\phi/L)
\end{equation}
where $k=2\pi n/L$.   Hence with two up fermions and two down fermions we can reproduce
(\ref{limit}) by having one fermion of each type with $k=0$ and $-\pi/2$.  

It turns out that the unprotected dimension 5 operator has an even simpler solution.  One starts by choosing a  basis of states with total spin $1/2$.  The possible states that can mix with each other are those invariant under shifts of the lattice sites.  The Hubbard Hamiltonian is also invariant under parity transformations that map site $i$ to $L-i$.  Since the states with all sites singly occupied are invariant under this transformation \cite{Beisert:2003tq}, we only need to consider states that are parity even.  This then leaves the following nine states that can mix with each other
\begin{eqnarray}\label{basis2}
\big| A\brangle &= &\frac{1}{\sqrt{2}} \Big(\big| \up\up\up\down\down\brangle_C-\big| \up\up\down\up\down\brangle\Big)\nonumber\\
\big| B_1\brangle&=&\frac{1}{2}\Big(\big| \updown0\down\up\up\brangle_C-\big| \updown0\up\up\down\brangle_C+\big| 0\updown\down\up\up\brangle_C-\big| 0\updown\up\up\down\brangle_C\Big)\nonumber\\
\big| B_2\brangle&=&\frac{1}{2\sqrt{3}}\Big(\big|  0\up\updown\down \up\brangle_C+\big|  0\up\updown \up\down \brangle_C-
2\big|  0\down\updown\up\up\brangle_C\nonumber\\
&&\qquad\qquad\qquad-\big|  \updown\up 0\down \up\brangle_C-\big| \updown\up 0 \up\down \brangle_C+
2\big|  \updown\down 0\up\up\brangle_C\Big)\nonumber\\
\big| B_3\brangle&=&\frac{1}{2}\Big(\big|  0\up\updown\down\up\brangle_C-\big|  0\up\updown \up\down \brangle_C+\big|  \updown\up 0\down\up\brangle_C-\big| \updown\up 0 \up\down \brangle_C\Big)
\nonumber\\
\big| B_4\brangle&=&\frac{1}{2\sqrt{3}}\Big(2\big|  \updown 0\up\down\up\brangle_C-\big|  \updown 0\down\up\up\brangle_C-\big|  \updown 0\up\up \down\brangle_C\nonumber\\
&&\qquad\qquad\qquad-2\big|  0\updown\up\down\up\brangle_C+\big|  0\updown\down\up\up\brangle_C+\big|  0\updown\up\up \down\brangle_C
\Big)\nonumber
\end{eqnarray}
\begin{eqnarray}
\big| C_1\brangle&=&\frac{1}{\sqrt{2}}\Big(\big|  \updown 0 \updown 0\up\brangle_C+\big| 0 \updown 0 \updown \up\brangle_C \ \Big)\nonumber\\
\big| C_2\brangle&=&\frac{1}{\sqrt{2}}\Big(\big|  \updown \updown 0 \up 0\brangle_C+\big| 0\, 0 \updown \up \updown\brangle_C \ \Big)\nonumber\\
\big| C_3\brangle&=&\frac{1}{\sqrt{2}}\Big(\big|  \updown \updown 0\, 0 \up\brangle_C+\big| 0\, 0 \updown \updown \up\brangle_C \ \Big)\nonumber\\
\big| C_4\brangle&=&\frac{1}{\sqrt{2}}\Big(\big|  \updown \updown 0 \up 0\brangle_C-\big| 0\, 0 \updown \up \updown\brangle_C \ \Big)\,,
\end{eqnarray}
where $\big| abcde\brangle_C$ refers to the cyclic state
\begin{equation}
\big| abcde\brangle_C=\frac{1}{\sqrt{5}}\Big(\big| abcde\brangle+\big| bcdea\brangle+\big| cdeab\brangle+\big| deabc\brangle+\big| eabcd\brangle\Big)
\end{equation}
Since the number of sites is odd, we can reproduce the BDS result with $\phi=0$ \cite{Rej:2005qt}.  Applying (\ref{ham}) to (\ref{basis2}), we find the Hamiltonian matrix to be
\begin{equation}\label{ham5}
H=\left(\begin{array}{ccccccccc}
5&2g&0&0&0&0&0&0&0\\
2g&4&\sqrt{\frac32}g&0&0&-g&-g&0&0\\
0&\sqrt{\frac32}g&4&\sqrt{\frac32}g&0&0&0&0&0\\
0&0&\sqrt{\frac32}g&4&\sqrt{\frac32}g&-g&0&-g&0\\
0&0&0&\sqrt{\frac32}g&4&0&0&0&-\sqrt{3}g\\
0&-g&0&-g&0&3&0&0&0\\
0&-g&0&0&0&0&3&0&0\\
0&0&0&-g&0&0&0&3&\sqrt{2}g\\
0&0&0&0&-\sqrt{3}g&0&0&\sqrt{2}g&3\\
\end{array}\right)
\end{equation}
Remarkably, the secular equation factorizes into a cubic and a sextic polynomial, with
\begin{eqnarray}\label{secular2}
\det(\Delta-H)&=&\frac{1}{4}\,P(\Delta)Q(\Delta)
\end{eqnarray}
where
\begin{eqnarray}
P(\Delta)&=&\Delta^3-11\Delta^2-10g^2\Delta+39\Delta+34g^2-45\nonumber\\
Q(\Delta)&=&4\Delta^6-88\Delta^5-30g^2\Delta^4+80\Delta^4+432g^2\Delta^3-3904\Delta^3+25g^4\Delta^2\nonumber\\
&&\qquad-2322g^2\Delta^2+10624\Delta^2-170g^4\Delta+5520g^2\Delta-15360\Delta\nonumber\\ &&\qquad+289g^4-4896g^2+9216\,.
\end{eqnarray}
The desired solution is a zero of the cubic, which can easily be solved as a series expansion in $g^2$,
given by
\begin{eqnarray}
\Delta&=&5+4 g^2-6 g^4+17 g^6-\frac{115 g^8}{2}+\frac{833 g^{10}}{4}-\frac{6147 g^{12}}{8}+\frac{44561g^{14}}{16}-\frac{303667 g^{16}}{32}
\nonumber\\
&&\qquad\qquad+\frac{1778945 g^{18}}{64}-\frac{6255235 g^{20}}{128}-\frac{45306735 g^{22}}{256}+\dots\,.
\end{eqnarray}
This reproduces the result in \cite{Rej:2005qt} using the Bethe ansatz.

We can also write down a simple closed form expression for $\lambda$ in terms of $\Delta$, where we find 
\begin{eqnarray}
\lambda=8\pi^2\frac{(\Delta^3-11\Delta^2+39\Delta-45)}{(10\Delta-34)}\,.
\end{eqnarray}
From this we then deduce that for large $\lambda$, $\Delta\approx\sqrt{5\lambda}\,/(2\pi)$, and so using the dispersion relation in (\ref{dispersion}),   the up spins have momenta $k=-2\pi/5,0,2\pi/5$ and the down spins have
$k=-2\pi/5,2\pi/5$.  Interestingly, for large $\lambda$ one sees that the $\Delta$ for this operator will be less than the $\Delta$ for the bare dimension 4 operator previously considered.  

The cubic $P(\Delta)$ is itself the secular equation for a reduced Hamiltonian given by
\begin{equation}
H_3=\left(\begin{array}{ccc}
5&2\sqrt{2}g&0\\
2\sqrt{2}g&3&\sqrt{2}g\\
0&\sqrt{2}g&3\\
\end{array}\right)\ ,
\end{equation}
while the sextic equation is the secular equation for
\begin{equation}
H_6=\left(
\begin{array}{llllll}
 4 & \sqrt{\frac{3}{2}} g & 0 & 0 & -g & 0 \\
 \sqrt{\frac{3}{2}} g & 4 & \sqrt{\frac{3}{2}} g & 0 & 0 & 0 \\
 0 & \sqrt{\frac{3}{2}} g & 4 & \sqrt{\frac{3}{2}} g & -g & -g \\
 0 & 0 & \sqrt{\frac{3}{2}} g & 4 & 0 & 0 \\
 -g & 0 & -g & 0 & 3 & 0 \\
 0 & 0 & -g & 0 & 0 & 3
\end{array}
\right)\ .
\end{equation}
The unitary transformation that rotates (\ref{ham5}) into the block form $H_3\oplus H_6$ is a complicated $g$ dependent transformation.  This   block diagonal form is probably due to the underlying integrability of the theory.

\sectiono{Large coupling for general states}

At large 't Hooft coupling $\lambda$, the Hubbard model essentially reduces to a theory of free lattice fermions.  Hence,  the energies of the Hubbard states in the leading order approximation are explicitly known.  What is not so trivial is determining which free fermion states go with the gauge invariant operators.   In this section we will study this problem for the operators with two impurities and then provide some of the details for more general operators.  Along the way, we will conjecture which  Hubbard states correspond to gauge invariant operators for any value of $\lambda$. 

  We assume that $L=2K+1$ to avoid any phase $\phi$.  It is known from the 1 loop results that if $M=2$ there are $K$ possible solutions for the gauge operators, so we should expect the same number of solutions to apply in the free fermion limit of large $\lambda$.  One of these is the chiral primary which has $\Delta=L$ and has all momentum modes singly occupied.  The  operators which are not chiral primaries will have $L-2$  different momentum modes for the up spins and  two different modes for the down spins, with an overlap of the up and down modes.  The sum of the  momenta  should be  a multiple of $2\pi$ in order for the state to have cyclic symmetry. 
 
The problem now is to determine which momentum modes are occupied.  To answer this, we consider the Bethe ansatz results in \cite{Rej:2005qt}.  The half-filled Hubbard model with $\phi=0$ has a nested Bethe ansatz whose equations are \cite{Lieb_Wu}
\begin{eqnarray}\label{bethe}
e^{iq_mL}&=&\prod_{j=1}^M \frac{u_j-\sqrt{2}g\sin q_m-i/2}{u_j-\sqrt{2}g\sin q_m+i/2}\,\,,\qquad m=0\dots L-1\nonumber\\
\prod_{m=0}^{L-1}\frac{u_j-\sqrt{2}g\sin q_m+i/2}{u_j-\sqrt{2}g\sin q_m-i/2}&=&\prod_{k\ne j}^M\frac{u_j-u_k+i}{u_j-u_k-i}\,,\qquad\qquad j=1\dots M.
\end{eqnarray}
The number of down spins is $M$ and the energy of the state is given by 
\begin{equation}
\Delta=L+\sqrt{2}g\sum_{m=1}^L \cos q_m\,.
\end{equation}
 In the limit where $g\to0$, the first Bethe equation implies
\begin{equation}
e^{iq_mL}=e^{iq_{m'}L}
\end{equation} for all $m$ and $m'$, while the second equation reduces to the Heisenberg chain Bethe equations
\begin{equation}
\prod_{m=1}^L\frac{u_j+i/2}{u_j-i/2}=\prod_{k\ne j}^M\frac{u_j-u_k+i}{u_j-u_k-i}\,.
\end{equation}
The values of $q_m$ that correspond to the gauge invariant operators in the $g\to0$ limit have
$q_m=2\pi m/L$ for $0\le m\le L-1$ \cite{Rej:2005qt},  and so
\begin{equation}\label{momH}
1=\prod_{j=1}^M \frac{u_j-i/2}{u_j+i/2}\,.  
\end{equation}

Let us consider the case where $M=2$. It was observed in  \cite{Rej:2005qt} that
the Bethe equations have a $Z_2$ symmetry under $q_m\to-q_m$, $u_j\to -u_j$.  In the Heisenberg limit,
(\ref{momH}) implies $u_1=-u_2\equiv u>0$, so the corresponding solutions will maintain the $Z_2$ symmetry for $g\ne0$.  Since $L$ is odd, this means that $q_0=0$ for all $g$ and $q_m=-q_{L-m}$. 
Hence the equations in (\ref{bethe}) reduce to
\begin{eqnarray}
e^{iq_mL}&=&\ \frac{u^2-(\sqrt{2}g\sin q_m+i/2)^2}{u^2-(\sqrt{2}g\sin q_m-i/2)^2}\equiv e^{i\chi_m}\,\,,\qquad\qquad m=1\dots K\label{phase1}\\
e^{2\pi in}&=&\prod_{m=1}^{K}\frac{(u+i/2)^2-2g^2\sin^2 q_m}{(u-i/2)^2-2g^2\sin^2 q_m}\equiv \prod_{m=1}^{K}e^{i\theta_m}\,.\label{phase2}
\end{eqnarray}
At $g=0$, we set the phases to $\chi_m=0$ and $\theta_m=2\pi n/K$,  for some $n$ satisfying $1\le n\le K/2$.   As $g$ is turned on,  $u$ must start increasing, otherwise all $\theta_m$ will start decreasing and  the equality in (\ref{phase2}) will not be maintained.    It is also clear that some $\theta_m$ must start decreasing toward 0 while others will start increasing toward $2\pi$, where those heading toward zero have the smaller values of $\sin q_m$.  In the limit of large $g$, it must happen that $u\sim g$, and that $n$ of the $\theta_m$ approach $2\pi$.

Examining (\ref{phase1}), one sees that $\chi_m$ approaches $0$ if $\theta_m$ approaches $0$,  since the real parts of the numerator and demonitor in (\ref{phase1})  differ from their counterparts in (\ref{phase2})  by a finite amount, while the imaginary parts of the numerators and denominators  do not change sign.   This means that $q_m$ flows back to $2\pi m/L$ when $g\to\infty$.  However, $\chi_m$ flows to $-2\pi$ if  $\theta_m$ flows to $2\pi$,  meaning that $q_m$ shifts to $2\pi(m-1)/L$.  If $q_m$ shifts and $q_{m-1}$ does not, then it must be that $u$ flows to $\sqrt{2}g\sin\frac{2\pi(m-1)}{L}$ in such a way that 
\begin{eqnarray}
u^2-2g^2\sin^2q_{m-1}>+1/4&&\nonumber\\
u^2-2g^2\sin^2q_{m}<-1/4&&\,.\nonumber
\end{eqnarray}

 In this analysis, we have assumed that $q_m$ is real throughout the flow. This is certainly true for $g$ small enough.   However, two $q_m$ can start moving into the complex plane once their real parts are equal.  But as we have just seen, $q_m$ can equal $q_{m-1}$ only in the limit $g\to\infty$, hence the $q_m$ remain real.

The largest values of $\sin q_m$ are centered around $m=K/2$.  In the flow, there will be $n$ sequential $q_m$ centered around $m=K/2$ that shift down to $2\pi(m-1)/L$.  The negative $q_m$  of this sequence also shift.
Hence, the final state for large $g$ has a two doubly occupied modes at $q=\pm\pi(K-n+\beta)/L$ and two unoccupied modes at $q=\pm\pi(K+n+\beta)/L$ where $\beta=(1-(-1)^{K+n})/2$.  (Figure 1 shows the flow for $L=17$ and $n=2$.)
Using perturbation theory about these free fermion states, the energies are found to be 
 \begin{equation}\label{LDel}
 \Delta= L+2\frac{\sqrt{\lambda}}{2\pi}\left(\cos\left(\frac{\pi(K-n+\beta }{L}\right)-\cos\left(\frac{\pi(K+n+\beta)}{L}\right)\right) - \frac{2(L-2)}{L}+{\rm O}\left(\frac{L}{\sqrt{\lambda}}\right)\,.
 \end{equation}
 
 Examining (\ref{LDel}), we see that $L/\sqrt{\lambda}$ is the effective expansion parameter and so (\ref{LDel}) is a good approximation if $L<<\sqrt{\lambda}$.  In fact, the convergence appears better as $n$ increases.  In any case, $L\sim\sqrt{\lambda}$ is where $\Delta$ crosses over from BMN  to free lattice fermion behavior.
     In the appendix we compute the first   order correction for all $n$ as well as the second order correction for $n<<L$.
       
\begin{figure} 
 \centering 
 \includegraphics[width=.9\textwidth]{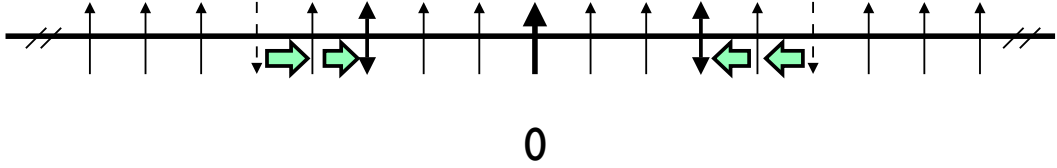} 
 \caption{A flow with $n=2$ for $M=2$ and $L=17$.  After the flow there are doubly occupied modes at $q=\pm6\pi/17$ and empty modes at $q=\pm10\pi/17$.} 
 \label{flow} 
\end{figure}

  In principle this analysis can be done for $M>2$.  For one thing, it appears that the equations in (\ref{bethe}) are consistent if  all $q_m$ are real, as long as for every $u_j$ there exists a $u_k$ such that $u_k=u_j^*$.  The $q_m$ then remain real throughout the flows and no two $q_m$ will be equal for finite $g$, preventing any flow into the complex plane. In fact, what seems to distinguish between Hubbard states that correspond to gauge invariant operators and those that do not are that the former have real $q_m$ with $\sum q_m=0\ {\rm mod}\ 2\pi$.   To see that the latter do not have both properties, note that in the limit $g\to 0$, a state with total momentum zero but not  corresponding to a gauge invariant operator will have $\Delta< L$.  The only way that this can occur is if $\cos q_m\sim g^{-1}$ for some $q_m$.  But this itself only happens if ${\rm Im}\, q_m\sim \pm\ln g$.

The $q_m$ can be thought of as momentum modes for charge density waves, while the $u_j$ are momentum modes for spin waves.  Real $q_m$  means that no two charge density modes are bound together.  In the $g\to0$ limit where the potential term dominates, this means that no two sites are doubly occupied.  

 If the $u_j$ are real, with $M_+$ of the $u_j$ positive  and $M_-$  negative, then our general expectation is that at least one of the positive $q_m$ shifts  by $-2\pi M_+/L$ and at least one of the negative $q_m$ shifts by $+2\pi M_-/L$.  It is also clear that if $q_m$ shifts by $-2\pi\ell/L$, then
 $q_{m-1}$ and $q_{m-2}$ must shift by at least $-2\pi(\ell-1)/L$ to avoid triple occupancy or modes hopping over each other.  In general, for the positive $q_m$ we will find a set of sequences, where the first is a sequence of $n_1$ modes shifting by $-2\pi/L$, followed by a seqence of $n_2$ shifting by $-4\pi/L$, {\it etc.}, all the way up to a sequence of $n_{M_+}$, where $n_i\ge 2$ for $i<M_+$.  Behind this last sequence, there can be sequences shifting by ever smaller amounts.  Figure 2 shows the example with $n_1=2$ and $n_2=1$ for $M=4$ and $L=17$.   If some of the $u_j$ are complex,
 then  it will be possible to have multiple sets of these sequences.  It is clear from this construction that
 the final configuration  will have $M$ of the momentum modes doubly occupied  and $M$ empty.  
 
  \begin{figure} 
 \centering 
 \includegraphics [width=0.9\textwidth]{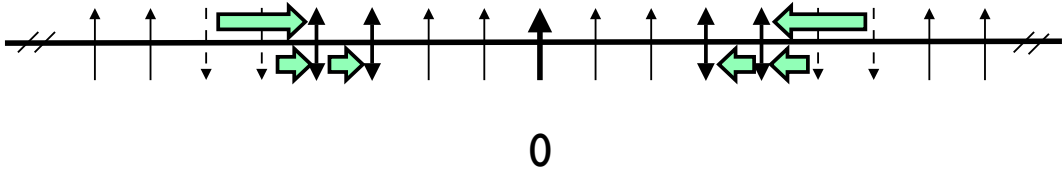} 
 \caption{A flow with $n_1=2$ and $n_2=1$ for $M_+=M_-=2$ and $L=17$.  After the flow there are doubly occupied modes at $q=\pm6\pi/17$ and $q=\pm8\pi/17$ and empty modes at $q=\pm10\pi/17$ and $q=\pm12\pi/17$.} 
 \label{flow2} 
\end{figure}

 As an example of this construction, consider the antiferromagnetic state ($M=K$) studied in \cite{Rej:2005qt,Zarembo:2005ur}.  This state would be expected to have all modes with $0<|k|<\pi/4$ to be doubly occupied and those with $\pi/4<|k|<\pi/2$ to be empty.  The sequences of shifts has $n_i=2$ for $i<K/2$.  The total energy of this state at strong coupling is
 \begin{equation}
\Delta_{AF}\approx 4\sqrt{2}g\sum_{n=1}^K\cos\frac{2\pi n}{L}\approx \frac{L\sqrt{\lambda}}{\pi^2}\,,
\qquad L>>1
\end{equation}
which agrees with the result in \cite{Zarembo:2005ur}.

\sectiono{Comparing strong coupling to string theory}
 
Let us now ask if the Hubbard model at large $\lambda$ can reproduce nonclassical string results.   At first sight this would seem problematic since generic string states
are expected to have energies satisfying
\begin{equation}\label{GKP}
\Delta^2-\vec p^{\,\,2}\approx4\tilde n\,\sqrt{\lambda}
\end{equation} 
for large $\lambda$, where $\tilde n$ is the level and $\vec p$ is a vector of Lorentz and $R$-charges  corresponding to the string center of mass motion \cite{Gubser:1998bc}.  This behavior can also be seen using a string Bethe ansatz \cite{Arutyunov:2004vx}.  But it seems from our discussion so far that $\Delta^2$ will be proportional to $\lambda$.  In fact
 we can see that for any length chain and for any generic state of that chain, $\Delta$ will scale as $\sqrt{\lambda}$ for large enough $\lambda$.       As was emphasized in \cite{Rej:2005qt}, the Hubbard model does not have order of limits ambiguities, so any discrepancy cannot be due to this.
 
 In any event, let us see how one could possibly reproduce the string results by considering the case $M=2$, and assume   that $L>>1$.  In this limit (\ref{LDel}) simplifies to
 \begin{equation}\label{Delapprox}
 \Delta\approx L+\frac{2n\sqrt{\lambda}}{L}-2\,
 \end{equation}
 for $n<<L$.
 The $R$-charge of our state is $J= L-2$ which corresponds to an angular momentum on $S^5$.  Hence we have
 \begin{equation}\label{main}
 \Delta^2-J^2\approx 4n\sqrt{\lambda}
 \end{equation}
 which agrees with (\ref{GKP}) if we identify $n$ with the level $\tilde n$.  Note that this approximation is valid
 as long as $\sqrt{\lambda}>>L>>\lambda^{1/4}$, which is outside the BMN limit and thus
 beyond the range of validity for the BDS Bethe equations.  Note that for a given $n$, if we reduce $\lambda$ such that we approach the BMN limit, then $n$ is the level of the string in the plane wave limit
 \cite{Berenstein:2002jq}.

For states with $M>2$,  the energies will still  have the form in
(\ref{LDel}), that is, as a sum over   doubly occupied modes minus a sum over  empty modes.  Hence, if the mode numbers are close enough to $K/2$ then $\Delta$ will still have the form in (\ref{Delapprox}) for some integer $n$.  However, it also seems that the  lattice fermion levels do not precisely match with with the BMN levels.
For example, for weak coupling and $M=4$ we can build the BMN like state where $u_1=-u_3\approx L/(2\pi)$ and $u_2=-u_4\approx L/(4\pi)$.  We would identify this state with a string state at level 3 \cite{Minahan:2002ve}.  However, for strong coupling this state would flow to a state with two consecutive doubly occupied modes followed by two empty modes.  Hence, this state has $n=4$.  It would be interesting to clarify this point.

If $L\sim\lambda^{1/4}$ then the approximation in (\ref{main}) starts breaking down.  If $L<<\lambda^{1/4}$, then $\Delta^2-J^2>>\sqrt{\lambda}$ for $\lambda>>1$.  One possibility is that the string dual is a highly excited state, with excitations along all directions of $AdS_5\times S^5$.  The Konishi operator would be one example of this\footnote{It has been  been shown that in order for the Konishi operator to have $\lambda^{1/4}$ behavior at strong coupling, it cannot be a rational function of $\sqrt{\lambda}$ \cite{Tseytlin:2004xa}.}.   It is also possible that these states are dual to semiclassical long strings \cite{Ryzhov:2004nz,Tseytlin:2004xa,Arutyunov:2004yx},
which typically scale as $\sqrt{\lambda}$.

There should be string states with small $R$-charges and squared masses of the same order as that in (\ref{GKP}), but presumably their corresponding operators are not in the $SU(2)$ sector.  But it also seems that these operators might not have small bare dimensions either.   Perhaps recent work on quantum integrability of the sigma model will lead to insights on this \cite{Mann:2005ab,Klose:2006dd,Gromov:2006dh}.   

 It is not clear how our results relate to the three loop discrepancy between gauge and classical string results \cite{Callan:2003xr,Serban:2004jf}.   It was pointed out in \cite{Minahan:2005jq} that the classical string solutions have small mixings outside of the $SU(2)$ sector that could account for the discrepancy.   This might require a generalization of the Hubbard model to include the entire operator spectrum in order to fully sort this out.  One might make progress here by generalizing the antiferromagnet solution of \cite{Rej:2005qt,Zarembo:2005ur}, as was recently done for the $SU(1,1)$ sector  in \cite{Arutyunov:2006av,SS} by using the asymptotic Bethe equations in \cite{Staudacher:2004tk,Beisert:2005fw}.   
  
\bigskip
\noindent {\bf Acknowledgments}:  This work was supported in part by Vetenskapr\aa det.

\renewcommand{\theequation}{A.\arabic{equation}}
 \setcounter{equation}{0}
\setcounter{section}{1} \setcounter{subsection}{0}
 \section*{Appendix A:  First and second order perturbative corrections}

We wish to find the first and second order perturbative corrections to states of the form
 $\big|n_\updown,-n_\updown;n_0,-n_0\brangle$, where $\pm n_\updown$ are  doubly occupied modes and $\pm n_0$ are empty modes.   If $n_\updown=n_0$ then the empty mode and the doubly occupied mode have combined to make singly occupied modes.
Defining the Fourier transform of $a_{\s,j}$ as 
\begin{equation}
\ta_{\s,n}=\frac{1}{\sqrt{L}}\sum_{j=1}^Le^{-2\pi i nj/L}\ta_{\s,j}\,,
\end{equation}
the four fermion interaction becomes
\begin{equation}
H_4=-\frac{1}{L}\sum_{n_1,n_2,n_3,n_4}\ta^\dagger_{\up,n_1}\ta^\dagger_{\down,n_2}\ta_{\up,n_3}\ta_{\down,n_4}\delta(n_1+n_2-n_3-n_4)\,.
\end{equation}
Hence for the desired states, the first order correction to the energy is
\begin{equation}
\Delta^{(1)}=\big\langle n_\updown,-n_\updown,n_0,-n_0\big|H_4|n_\updown,-n_\updown,n_0,-n_0\big\rangle
=-\frac{2(L-2)}{L}\,.
\end{equation}

For the second order correction, we assume that $n_\updown=(K-n+\beta)/2$ and $n_0=(K+n+\beta)/2$, where $\beta$ is defined in the text.  In general, we have the following matrix elements
\begin{eqnarray}\label{matelems}
&&\big\langle n_\updown,-n_\updown,n_0,-n_0\big|H_4\big|n_\updown+m,-n_\updown;n_0+m,-n_0\brangle=-1/L\nonumber\\
&&\big\langle n_\updown,-n_\updown,n_0,-n_0\big|H_4\big|n_\updown,-n_\updown+m;n_0,-n_0+m\brangle=-1/L\nonumber\\
&&\big\langle n_\updown,-n_\updown,n_0,-n_0\big|H_4\big|n_\updown+m,-n_\updown;n_0,-n_0+m\brangle=-1/L\nonumber\\
&&\big\langle n_\updown,-n_\updown,n_0,-n_0\big|H_4\big|n_\updown,-n_\updown+m;n_0+m,-n_0\brangle=-1/L\,.
\end{eqnarray}
We also have that the difference in energies are
\begin{eqnarray}\label{endiff1}
&&\Delta^{(0)}_{n_\updown,-n_\updown,n_0,-n_0}-\Delta^{(0)}_{n_\updown+m,-n_\updown;n_0+m,-n_0}
\nonumber\\
&&\qquad=\frac{\sqrt{\lambda}}{2\pi}\Bigg[\cos\frac{\pi(K-n+\beta)}{2K+1}-\cos\frac{\pi(K+n+\beta)}{2K+1}\nonumber\\
&&\qquad\qquad\ \ -\cos\frac{\pi(K-n+2m+\beta)}{2K+1}+\cos\frac{\pi(K-n+2m+\beta)}{2K+1}\Bigg]\nonumber\\
&&\qquad\approx\frac{n\pi^2\sqrt{\lambda}}{L^3}(2m^2+(1-2\beta)m)\,,
\end{eqnarray}
\begin{eqnarray}\label{endiff2}
&&\Delta^{(0)}_{n_\updown,-n_\updown,n_0,-n_0}-\Delta^{(0)}_{n_\updown+m,-n_\updown;n_0,-n_0+m}
\nonumber\\
&&\qquad=\frac{\sqrt{\lambda}}{2\pi}\Bigg[\cos\frac{\pi(K-n+\beta)}{2K+1}-\cos\frac{\pi(K+n+\beta)}{2K+1}\nonumber\\
&&\qquad\qquad\ \ -\cos\frac{\pi(K-n+2m+\beta)}{2K+1}+\cos\frac{\pi(K-n+2m+\beta)}{2K+1}\Bigg]\nonumber\\
&&\qquad\approx-\frac{\sqrt{\lambda}}{L}(2m)\,.
\end{eqnarray}
Since (\ref{endiff1}) is large compared to (\ref{endiff2}), we can ignore the contributions coming from the $\big|n_\updown+m,-n_\updown;n_0,-n_0+m\brangle$ and $\big|n_\updown,-n_\updown+m;n_0+m,-n_0\brangle$ states.  We then find,
\begin{eqnarray}
\Delta^{(2)}&\approx&2\sum_{m\ne 0}\frac{L^3(-1/L)^2}{n\pi^2\sqrt{\lambda}(2m^2+(1-2\beta)m)}=
\frac{4L}{n\pi^2\sqrt{\lambda}}\,.
\end{eqnarray}

\end{document}